\documentclass[times, review, 10pt]{elsarticle}

\usepackage{amssymb}
\usepackage{amsmath}

\usepackage{url}
\usepackage[T1]{fontenc}
\biboptions{numbers,sort&compress}
\usepackage{float}
\usepackage{booktabs}
\usepackage{multirow} 
\usepackage[symbol]{footmisc}
\usepackage{fontawesome}

\usepackage{changes}
\definechangesauthor[color=blue]{author}

\journal{NeuroImage}

\begin{document}

\begin{frontmatter}



\title{\textit{DeepNuParc}: A Novel Deep Clustering Framework for Fine-scale Parcellation of Brain Nuclei Using Diffusion MRI Tractography}


\author[1,2]{Haolin He}
\author[1]{Ce Zhu}
\author[1]{Le Zhang}
\author[1]{Yipeng Liu}
\author[1]{Xiao Xu}
\author[3]{Yuqian Chen}
\author[3]{Leo Zekelman}
\author[4]{Jarrett Rushmore}
\author[3]{Yogesh Rathi}
\author[3]{Nikos Makris}
\author[3]{Lauren J. O'Donnell\corref{senior}}
\author[1]{Fan Zhang\corref{senior}}

\cortext[senior]{Lauren J. O'Donnell and Fan Zhang contributed equally as co-senior authors. The corresponding author is Fan Zhang (fan.zhang@uestc.edu.cn).}

\affiliation[1]{organization={University of Electronic Science and Technology of China},
                city={Chengdu},
                country={China}}

\affiliation[2]{organization={The Chinese University of Hong Kong},
                city={Hong Kong SAR},
                country={China}}

\affiliation[3]{organization={Harvard Medical School},
                city={Boston},
                country={USA}}

\affiliation[4]{organization={Boston University},
                city={Boston},
                country={USA}}
\begin{abstract}
Brain nuclei are clusters of anatomically distinct neurons that serve as important hubs for processing and relaying information in various neural circuits.
Fine-scale parcellation of the brain nuclei is vital for a comprehensive understanding of their anatomico-functional correlations.
Diffusion MRI tractography is an advanced imaging technique that can estimate the brain's white matter structural connectivity to potentially reveal the topography of the nuclei of interest for studying their subdivisions.
In this work, we present a deep clustering pipeline, namely \textit{DeepNuParc}, to perform automated, fine-scale parcellation of brain nuclei using diffusion MRI tractography.
First, we incorporate a newly proposed deep learning approach to enable accurate segmentation of the nuclei of interest directly on the dMRI data.
Next, we design a novel streamline clustering-based structural connectivity feature for a robust representation of voxels within the nuclei.
Finally, we improve the popular joint dimensionality reduction and k-means clustering approach to enable nuclei parcellation at a finer scale.
We demonstrate \textit{DeepNuParc} on two important brain structures, i.e. the amygdala and the thalamus, that are known to have multiple anatomically and functionally distinct nucleus subdivisions.
Experimental results show that \textit{DeepNuParc} enables consistent parcellation of the nuclei into multiple parcels across multiple subjects and achieves good correspondence with the widely used coarse-scale atlases.
Our code is available at \url{https://github.com/HarlandZZC/deep_nuclei_parcellation}.
\end{abstract}




\end{frontmatter}


\section{Introduction}
\label{Introduction}
Brain nuclei are clusters of anatomically distinct neurons that serve as important hubs for processing and relaying information in various neural circuits \cite{vanHuijzen2009, Rushmore2022}.
For example, the amygdala is a collection of nuclei that is responsible for human emotional processing \cite{Puccetti2021}, and the thalamus contains various nuclei that act as relay stations for sensory information headed to the cerebral cortex \cite{Jones2012}.
Anatomically, these brain nucleus structures have a complex internal composition with several types of neurons arranged into multiple subdivisions, and each subdivision exhibits unique functional roles.
They are interconnected with other brain regions through white matter (WM) pathways that allow for the integration and coordination of various brain functions.
Many studies have shown that the distinct functional specificity of the subdivisions of these brain nucleus structures plays an important role in the progression and diagnosis of related brain diseases \cite{Alkemade2013, Nikolenko2020, Hwang2017, Herrero2002}.
For instance, abnormalities in specific amygdala subdivisions have been linked to brain disorders such as autism, Alzheimer's disease, and anxiety \cite{Nikolenko2020}.
Therefore, parcellation of brain nuclei is of great interest for understanding their neuroanatomical features and can potentially assist in the diagnosis and therapy of brain diseases.

Magnetic resonance imaging (MRI) is a widely used technique to study brain nucleus structures and their subdivisions.
The most commonly used is the structural MRI (sMRI), e.g., T1-weighted images.
In the past few decades, a large number of studies have used sMRI to investigate the anatomy of various brain nuclei and largely enhanced our understanding of their overall structure and function \cite{Richardson2009, Ries2008, Brunenberg2011}.
More recently, to enable investigation of the detailed structure and subdivisions, studies have utilized advanced high-resolution sMRI scans for improved imaging of the brain nuclei \cite{Saygin2017, Tyszka2016, Saranathan2021, Massey2012, Rushmore2020}.
For instance, nine amygdala nucleus subdivisions are delineated using postmortem specimens at high resolution (100–150 µm) on a 7T scanner \cite{Saygin2017}.
However, one shortcoming of sMRI is that it does not provide information about the inter-regional connectivity of the brain.
As a result, only using sMRI is insufficient in assessing how the nucleus structures of interest interact with other brain regions — information critical in assisting in identifying anatomically distinct subdivisions within the nuclei.

Moving beyond the conventional sMRI data, there is an increasing interest in developing connectivity-based nucleus parcellation methods using more advanced MRI techniques \cite{Eickhoff2015, Messe2020}.
One category of methods utilizes functional MRI (fMRI) to assess the functional connectivity between various brain regions \cite{Zhang2018a, Bielski2021, Yang2016, Moghimi2022}.
These methods parcellate the nucleus structures of interest into subdivisions with similar blood-oxygen-level-dependent (BOLD) signals, based on the assumption that voxels exhibiting similar activity share the same functional role.
For example, one recent fMRI study \cite{Bielski2021} subdivides the amygdala into two subdivisions and shows good correspondence with the widely used parcellation atlas that includes the laterobasal (LB), the centromedial (CM), and the superficial (SF) subdivisions \cite{Eickhoff2005}.
Another category of work attempts to use diffusion MRI (dMRI) for brain parcellation based on WM structural connectivity \cite{Wen2016, Avecillas2023, Saygin2011, Bisecco2015}, which aligns with the scope of our study.
dMRI is an advanced MR technique that uniquely enables in vivo reconstruction of the brain’s WM connections via a computational process called tractography \cite{Basser2000, Zhang2022}.
Many methods have been proposed for nucleus parcellation using dMRI tractography, under the assumption that voxels connected to the same brain regions share the same WM structural connectivity \cite{Wen2016, Avecillas2023, Saygin2011, Bisecco2015}. That is to say, voxels crossed by fibers with similar white matter anatomy or pathway trajectories typically belong to the same anatomical subdivision.
Such methods have demonstrated successful applications in studying brain diseases such as schizophrenia \cite{Basile2020} and Parkinson's disease \cite{ Zolal2020, Milardi2022}.

Despite the increasing popularity of connectivity-based nucleus parcellation, effective and fine-scale parcellation using dMRI tractography remains a challenge.
In general, the overall process for parcellating the nucleus structure of interest (e.g., the amygdala) into multiple subdivisions includes the following major steps:
1) segmentation, that is identifying the entire structure from the dMRI data,
2) connectivity feature extraction, that is computing imaging features that are informative to differentiate voxels belonging to different subdivisions, and
3) nucleus voxel clustering, that is grouping the voxels based on the extracted connectivity features.
There are several technical challenges during this process.
The first is how to accurately segment the nucleus structure of interest in dMRI data.
The typical approach involves segmenting structures on higher-resolution sMRI with superior tissue contrast, registering to dMRI space, and optionally refining the segmentation using dMRI-derived parameters to minimize partial volume contamination \cite{Najdenovska2018}.
However, the registration is challenging due to image distortions and low resolution of dMRI data, often resulting in segmentation errors between the nucleus of interest and its neighboring regions \cite{Zhang2023}.
The second challenge is how to come up with an informative connectivity feature representation that can discriminatively describe the subdivisions of the nucleus structure.
Existing work usually represents voxels using local fiber orientation distribution \cite{Wen2016} or connectivity probability derived from probabilistic tractography \cite{Avecillas2023, Saygin2011}.
However, these methods do not consider the actual trajectory of the WM fibers that reflect the underlying WM anatomy.
The third challenge is how to effectively identify the subregions of the nucleus structure of interest based on their connectivity features.
This has been done by using traditional machine learning clustering approaches, e.g., k-means applied to custom-designed voxel features \cite{Wen2016, Avecillas2023, Najdenovska2018}, to identify voxels with similar connectivity patterns, while recent advances in deep clustering have shown superior performance over traditional methods in dMRI-related parcellation tasks \cite{Chen2023, Xu2023}.

In this paper, we propose a novel deep-learning method, namely \textit{DeepNuParc}, for automated, fine-scale nucleus parcellation using dMRI tractography.
We demonstrate our method on two important brain nucleus structures, i.e. the amygdala and the thalamus, that are known to have multiple anatomically and functionally distinct nucleus subdivisions.
Fine-scale parcellation of these two nucleus structures is highly important in clinical and scientific applications, e.g., identifying targets for deep brain stimulation (DBS) in the treatment of neurodegenerative conditions.
Overall, \textit{DeepNuParc} includes three major technical contributions.
First, we segment the nucleus structure of interest directly from dMRI data to avoid any potential segmentation errors caused by inter-modality registration between sMRI and dMRI data.
Second, we design a novel streamline clustering-based connectivity feature to enable robust representation of the voxels within the segmented nucleus structure of interest.
Third, we propose a deep clustering network that extends the popular joint dimensionality reduction and k-means clustering approach \cite{Yang2017} to group voxels with similar connectivity for nucleus parcellation at a finer scale.
We perform experiments on the Human Connectome Young Healthy Adult (HCP-YA) dataset \cite{Stephen2013}.
Experimental results show that \textit{DeepNuParc} enables consistent nucleus structure parcellation into multiple subdivisions across multiple subjects and achieves good correspondence with the widely used coarse-scale atlases.
This investigation extends our previous conference publication \cite{He2024} to include an improved neural network training framework, additional experiments on a broader spectrum of nuclei (including the thalamus and the amygdala), a more comprehensive performance assessment, and an improved method description with in-depth technical details.

The rest of this paper is organized as follows.
Section \ref{Methods} describes the experimental datasets, the proposed framework, and the model training and testing.
Section \ref{Experiments and Results} presents the experimental setup and results on high-quality dMRI data.
Finally, the discussion, conclusions, and future work are given in Section \ref{Discussion and Conclusion}.

\begin{figure}[H]
\centering
\includegraphics[width=1\linewidth]{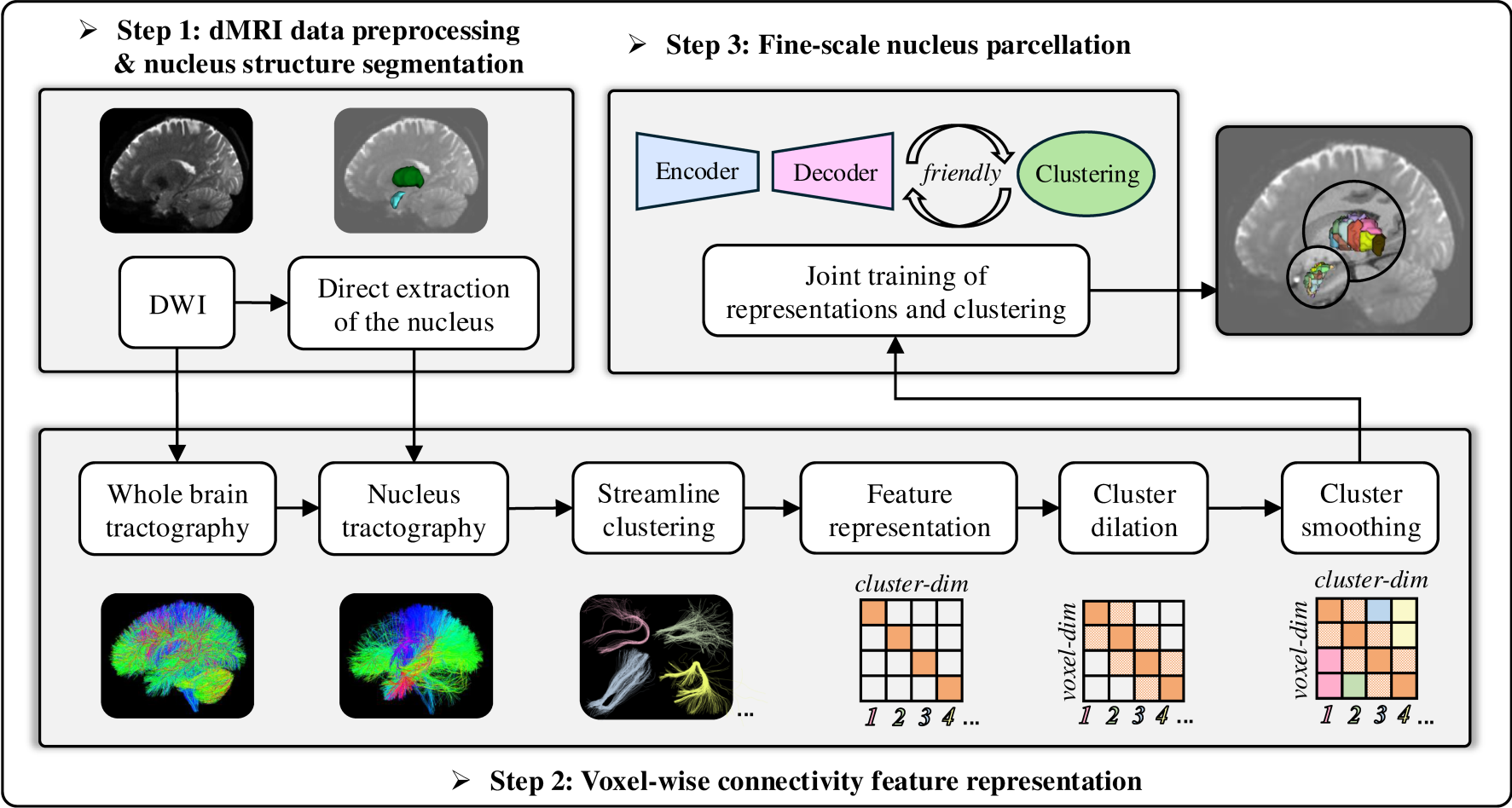}
\caption{Overall framework of \textit{DeepNuParc}.}\label{Overall framework of DeepNuParc.}
\end{figure}

\section{Methods}
\label{Methods}

Figure \ref{Overall framework of DeepNuParc.} gives an overview of our proposed \textit{DeepNuParc} method, including 
(1) dMRI data preprocessing and nucleus structure segmentation (Section \ref{Dataset, Preprocessing, and Nucleus Structure Segmentation}), 
(2) voxel-wise connectivity feature representation that leverages dMRI tractography streamline clustering (Section \ref{Nucleus Tractography and Streamline Clustering}) and streamline cluster dilation and smoothing (Section \ref{Feature Extraction via Streamline Cluster Dilation and Smoothing}), and 
(3) fine-scale nucleus parcellation using the proposed adaptive k-means-friendly autoencoder clustering method.

\subsection{Dataset, Preprocessing, and Nucleus Structure Segmentation}
\label{Dataset, Preprocessing, and Nucleus Structure Segmentation}

We utilize dMRI data from 100 young healthy adults (29.1±3.7 years; 54 females and 46 males) from the Human Connectome Project (HCP) \cite{Stephen2013}.
The acquisition parameters are TE=89.5ms, TR=5520ms, and voxel size=1.25×1.25×1.25 mm\textsuperscript{3}, 18 baseline images, and 270 diffusion-weighted images distributed evenly at b=1000/2000/3000 s/mm\textsuperscript{2}.
The provided dMRI data in HCP has been processed following the HCP minimum processing pipeline, including brain masking, motion correction, eddy current correction, EPI distortion correction, and rigid registration to the MNI space \cite{Glasser2013}.
In our experiments, 80 subjects are used for model training and 20 subjects are used for testing.

Segmentation of the entire nucleus structure (i.e., the amygdala and the thalamus) is performed using our recently proposed \textit{DDParcel} method \cite{Zhang2023} that performs the FreeSurfer Desikan-Killiany (DK) parcellation \cite{Desikan2006} including the nuclei of interest.
Unlike existing methods that perform segmentation on T1-weighted data and then register to dMRI space, \textit{DDParcel} enables segmentation directly from the dMRI data to avoid potential errors due to inter-modality registration.
In this way, we can obtain an accurate nucleus structure segmentation to benefit the following connectivity-based parcellation into multiple subdivisions.
In brief, the input of \textit{DDParcel} is the subject-specific dMRI-derived parameter maps including fractional anisotropy, mean diffusivity, and diffusion tensor eigenvalues.
A pre-trained segmentation model provided in \textit{DDParcel} with 101 anatomical regions (including both cortical and subcortical parcellations) corresponding to the DK parcellation is applied to the input parameter maps.
One benefit of \textit{DDParcel} is the use of a multi-level fusion network to leverage the complementary information from the multiple dMRI-derived parameters for accurate brain segmentation in dMRI space.

\subsection{Voxel-wise Connectivity Feature Representation}
\label{Voxel-wise Connectivity Feature Representation}

After segmentation of the entire nucleus structure of interest, we compute a structural connectivity feature representation of each voxel within the structure based on its WM connections.
Unlike existing work that represents voxels using local fiber orientation distributions \cite{Wen2016} or connectivity probabilities derived from probabilistic tractography \cite{Avecillas2023, Saygin2011}, we propose to use deterministic tractography data parcellated into streamline clusters for a robust connectivity representation, as deterministic tractography can provide highly geometrically plausible trajectory information and ensures the effectiveness of our subsequent streamline cluster dilation and smoothing processes (Section \ref{Feature Extraction via Streamline Cluster Dilation and Smoothing}).

\subsubsection{Nucleus Tractography and Streamline Clustering}
\label{Nucleus Tractography and Streamline Clustering}

We first compute tractography to extract all WM streamlines connecting to the nucleus structure of interest, which is referred to as “nucleus tractography” in our study.
To do so, we perform whole-brain tractography, followed by selecting all streamlines connected to the nucleus structure.
We use a dual-tensor Unscented Kalman Filter (UKF) method \cite{Reddy2016} provided in the SlicerDMRI software \cite{Norton2017, Zhang2020} to compute the whole-brain tractography.
We choose the UKF method because it has been demonstrated to be highly robust for successful fiber tracking across the lifespan \cite{Zhang2018b} and highly sensitive, especially in the presence of crossing fibers in WM structures \cite{Baumgartner2012, Chen2016, Liao2017, Xue2023}.
During the streamline filtering process, to avoid missing any potential streamlines that terminate before reaching the nucleus structure, we dilate the nucleus segmentation and use the dilated region as an inclusion mask to retain streamlines that pass through the nucleus from the whole-brain tractography.

Next, we perform streamline clustering to subdivide the nucleus tractography into multiple distinct fiber bundles.
Due to the large number of streamlines in nucleus tractography, it is difficult to build a feature representation directly for each voxel using individual streamlines.
Therefore, we perform streamline clustering (also known as fiber clustering) to group white matter streamlines based on their spatial and geometric properties.
We use our well-established WhiteMatterAnalysis (WMA) pipeline \cite{Zhang2018b, ODonnell2007, ODonnell2012} for groupwise streamline clustering simultaneously across multiple subjects.
In previous work, we have shown successful application of this method in whole-brain and region-specific tractography clustering analyses \cite{Zhang2018b, Xue2023, Zhang2020a}.
In our study, for each subject, we subdivide the nucleus tractography into \( K \) streamline clusters (see Section \ref{Determination of Parcel Quantity: Effect of Major Parameters} for parameter setting of \( K \)).

\begin{figure}[H]
\centering
\includegraphics[width=0.8\linewidth]{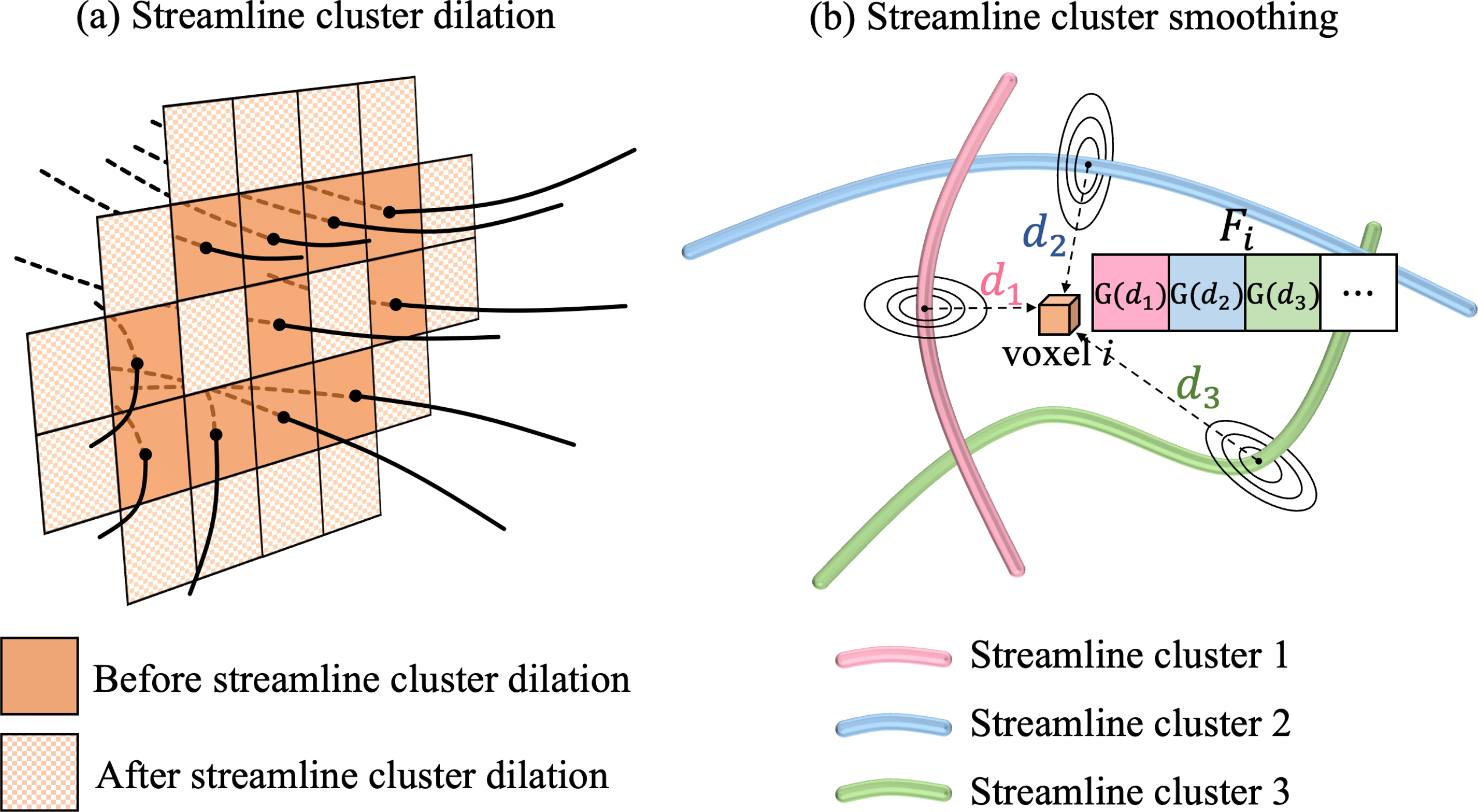}
\caption{Illustration of the streamline cluster dilation and smoothing.}\label{Illustration of the streamline cluster dilation and smoothing.}
\end{figure}

\subsubsection{Feature Extraction via Streamline Cluster Dilation and Smoothing}
\label{Feature Extraction via Streamline Cluster Dilation and Smoothing}
We then extract a feature representation for each voxel within the segmented nucleus structure per subject based on the obtained streamline clusters.
Based on our previous work for dentate parcellation \cite{Xu2023}, for each voxel $i$, we construct a feature representation $F_i = (v_{ij} \mid j = 1, \dots, K)$, where $v_{ij}$ is set to 1 if streamline cluster $j$ intersects with $i$.
However, this representation can be ineffective for brain structures like the amygdala and the thalamus which have highly complex WM connections.
First, there are voxels within the nucleus not directly intersecting with any streamline clusters (referred to as empty voxels), resulting in $F_i$ being a zero vector.
Second, this representation neglects the spatial relationship between the voxels and the clusters; that is to say, $v_{ij}$ is 0 no matter how far it is between $i$ and $j$.
To resolve these, we propose two novel additions for an improved feature representation.

First, to address empty voxels, we design a streamline cluster dilation process, as illustrated in Figure \ref{Illustration of the streamline cluster dilation and smoothing.}.a.
For each empty voxel $i$, we set $v_{ij} = 1$ if any adjacent voxel of $i$ intersects with cluster $j$.
This process largely reduces the sparsity effect of streamlines within a cluster, which results in voxels that are generally traversed by the streamline cluster but do not intersect with any individual streamlines.

Second, to provide information about spatial relationships, we design a streamline cluster smoothing process, as illustrated in Figure \ref{Illustration of the streamline cluster dilation and smoothing.}.b.
For each voxel $i$ with $v_{ij} = 0$, we set $v_{ij} = G(d_{ij})$, where $d_{ij}$ is the smallest distance between $i$ and all voxels intersecting with cluster $j$, and $G(d)$ is computed via a Gaussian kernel with mean $= 0$ and $\sigma = 1$.
After this, each element in $F_i$ provides information about the spatial relationship between the voxel and the streamline clusters, while ensuring no empty elements with a zero feature representation.

\subsection{Fine-Scale Nucleus Parcellation via Deep Clustering}
\label{Fine-Scale Nucleus Parcellation via Deep Clustering}

We parcellate the nucleus structure of interest by clustering the voxels based on their connectivity features.
To do this, we design a dense autoencoder network that extends the deep clustering method in \cite{Yang2017} for an improved fine-scale parcellation.
The overall network architecture is shown in Figure \ref{Illustration of the streamline cluster dilation and smoothing.}, including three major parts:
(1) encoder to extract a low-dimensional latent feature for each voxel based on its input connectivity feature,
(2) decoder to reconstruct the original input from the low-dimensional feature, and
(3) k-means to cluster voxels belonging to the segmented nucleus structure, thereby achieving nucleus parcellation into multiple parcels.

\begin{figure}[H]
\centering
\includegraphics[width=0.8\linewidth]{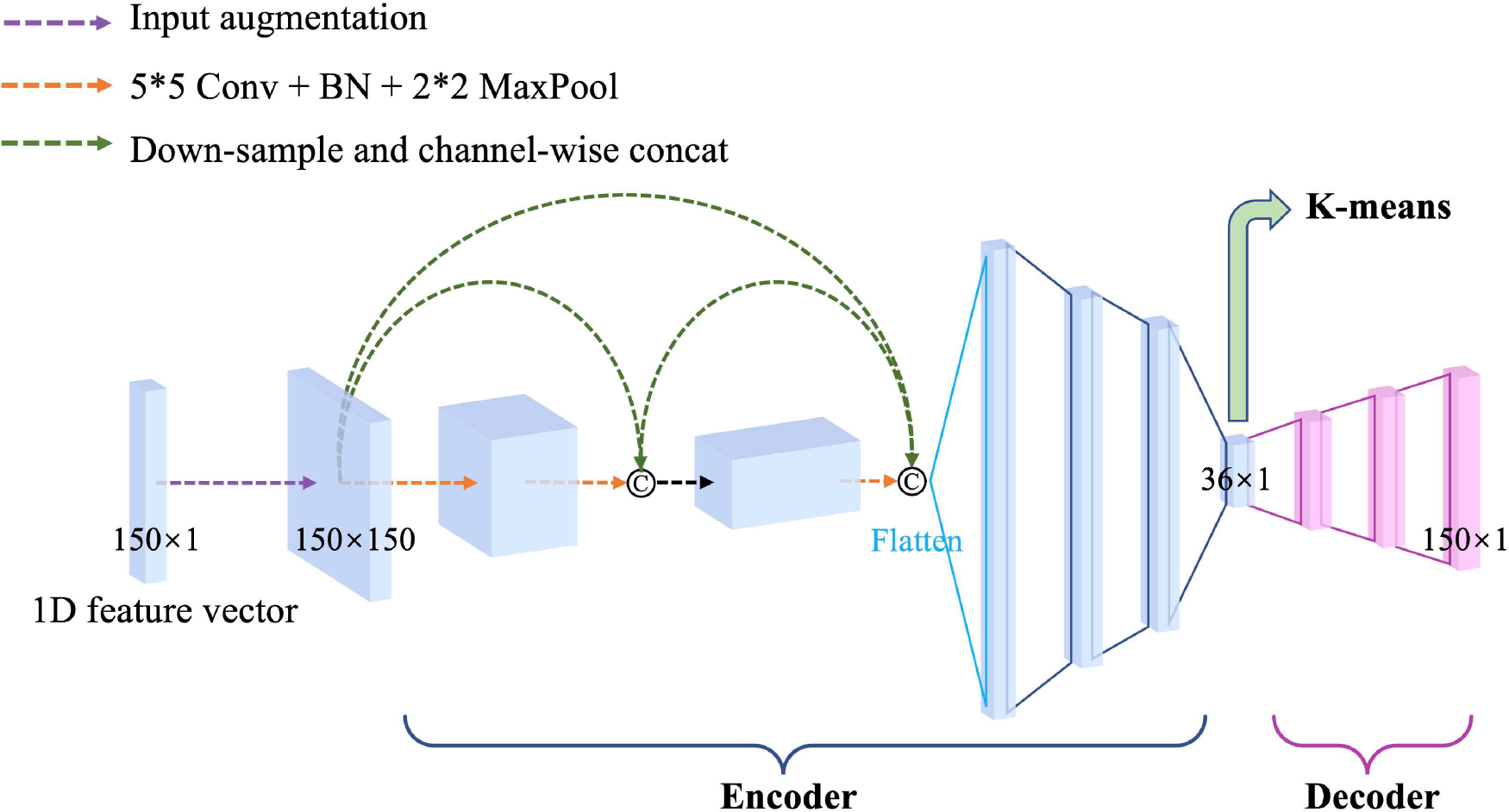}
\caption{Overview of the proposed adaptive k-means-friendly autoencoder clustering network.}\label{Overview of the proposed adaptive k-means-friendly autoencoder clustering network.}
\end{figure}

\subsubsection{Input Augmentation and Dense Autoencoder Framework}
\label{Input Augmentation and Dense Autoencoder Framework}

Compared to the original network proposed in \cite{Yang2017}, we first include an input augmentation process to address the specialty of the input feature.
Specifically, the original input is a 1D feature representation, where each element corresponds to one streamline cluster and its neighboring elements are two other random streamline clusters.
A straightforward way is to apply a 1D convolution network, but it can only use the randomly assigned neighborhood information.
The relationships between individual elements within each one-dimensional feature vector are critically important.
For instance, when two elements are both `1', it indicates that the nucleus of interest is simultaneously traversed by two streamline clusters.
However, due to the limited receptive field, 1D convolution struggles to capture each element equally.
Relationships between closely spaced elements are more readily captured by 1D convolution, while relationships between elements that are further apart are more challenging to capture.

Hence, similar to our previous work that builds a 2D representation from a 1D vector for tractography parcellation \cite{Zhang2020b}, we perform an input augmentation by shuffling the elements in the feature vector sequentially and concatenating them into a 2D input matrix.
In our study, we cyclically shifted the feature vector $K$ (the number of streamline clusters) times, generating a feature matrix with a size of $K \times K$.
Thus, we can apply a 2D convolution with a larger receptive field to representation learning.
Additionally, thanks to the cyclic arrangement of the elements, the 2D convolutional kernels can convolve the relationships between all elements in a uniform and equitable manner.

The CNN component consists of a $5\times5$ convolutional layer followed by batch normalization (BN) and $2\times2$ max pooling (as shown by the orange dashed arrows), which performs feature extraction and spatial dimension reduction at each encoder level.
Furthermore, we also improve the original simple convolution network with a DenseNet structure \cite{Huang2017} in the encoder to better use the input feature, as illustrated with the green dashed arrows in Figure \ref{Overview of the proposed adaptive k-means-friendly autoencoder clustering network.}.
Specifically, for the encoder, we incorporate a dense structure.
The output of each layer, after being downsampled, is concatenated with the output of the subsequent layer.
This architecture, combining dense and convolutional layers, can enable the encoder to better utilize the information from each convolutional layer, thereby capturing details from the feature representation.

\subsubsection{Adaptive K-Means-Friendly Training Mechanism}
\label{Adaptive K-Means-Friendly Training Mechanism}

One critical challenge for the deep clustering model is to ensure that the learned latent features from pre-trained autoencoders can be adapted according to the downstream clustering task.
While the joint dimensionality reduction and k-means clustering approach in \cite{Yang2017} has largely addressed this, we find it ineffective when a fine-scale clustering is performed, generating unbalanced clustering results where a subset of clusters is empty or with a small number of voxels.
Therefore, we propose an improved adaptive training mechanism to avoid this issue, as follows.

We first pre-train the autoencoder using a mean squared error (MSE) loss to learn a latent feature per voxel.
Then, we simultaneously train the autoencoder and the k-means clustering based on the latent feature computed from the pre-training stage.
This allows a fine-tuning of the autoencoder together with the clustering.
During each training batch, we check if there are any clusters with only a small number of voxels (which is empirically set to be fewer than $1/80$ of the batch size), and if so, replace these clusters’ centroids with the mean of the others.

Below is the loss function including one component for the autoencoder and one for cluster centroid assignment:
\[
Loss_{train} = \lambda \text{MSE}(d(e(x_i)), x_i) + \beta \|e(x_i) - m_{k \leftarrow i} \|_2^2,
\]

where $e(\cdot)$ is the encoder, $d(\cdot)$ is the decoder, $x_i$ is the input feature, and $m_{k \leftarrow i}$ is the cluster centroid to which $x_i$ belongs.
The weighting parameter $\lambda$ is set to be $15000$ based on a testing range from $5000$ to $20000$, and $\beta$ is set to be $0.5$ based on a testing range from $0.1$ to $10$.

\section{Experiments and Results}
\label{Experiments and Results}

To verify the reliability and generalizability of our method, we perform the following experimental evaluations on both the amygdala and thalamus.
First, we assess the influence of major parameters involved in our method (Section \ref{Determination of Parcel Quantity: Effect of Major Parameters}).
Second, we compare with several baseline methods to demonstrate the performance of each component in our proposed method (Section \ref{Comparison across Different Methods}).
Third, we compare our parcellation results with existing atlases to assess if our parcellation corresponds to the known anatomy of the nucleus structures of interest.

\subsection{Evaluation Metrics}
\label{Evaluation Metrics}

For exploratory parcellation tasks based on tractography in medical imaging, it is crucial to use appropriate metrics to evaluate the effectiveness of parcellation.
In our study, we use a total of four metrics, which are as follows:
1) \textit{Spatial Continuity (SC)} is the percentage of voxels in the maximum connected component per parcel, where a higher SC indicates a better spatial continuity.
2) \textit{Dice Coefficient (Dice)} measures the spatial overlap of the corresponding parcels across subjects, where a high value indicates a high consistency.
3) \textit{Relative Standard Deviation (RSD)} is calculated by taking the variance of the Fractional Anisotropy (FA) values of all voxels within each parcel, dividing it by the mean, and then averaging this result across all parcels, where a lower RSD indicates better distinctiveness between different parcels.

\subsection{Determination of Parcel Quantity: Effect of Major Parameters}
\label{Determination of Parcel Quantity: Effect of Major Parameters}

In our parcellation algorithm, two parameters are particularly important.
One is k, which represents the number of streamline clusters for clustering all streamlines passing through the nucleus structure.
The other is c, which represents the final number of parcels into which the target nucleus is divided.
To explore reasonable values for k and c, we investigate all hyperparameter combinations of k = 25, 50, 100, 150, 200, and c ranging from 3 to 13 for both the amygdala and thalamus.
For the parcellation results of each hyperparameter combination, we evaluate the effectiveness using the aforementioned three metrics: \textit{Dice}, \textit{SC}, and \textit{RSD}.
For \textit{SC} and \textit{RSD}, we compute the mean value across all testing subjects.
For \textit{Dice}, we compute the mean score across all testing subject pairs for each parcel and then the average \textit{Dice} across all parcels.

\begin{figure}[H]
\centering
\includegraphics[width=0.55\linewidth]{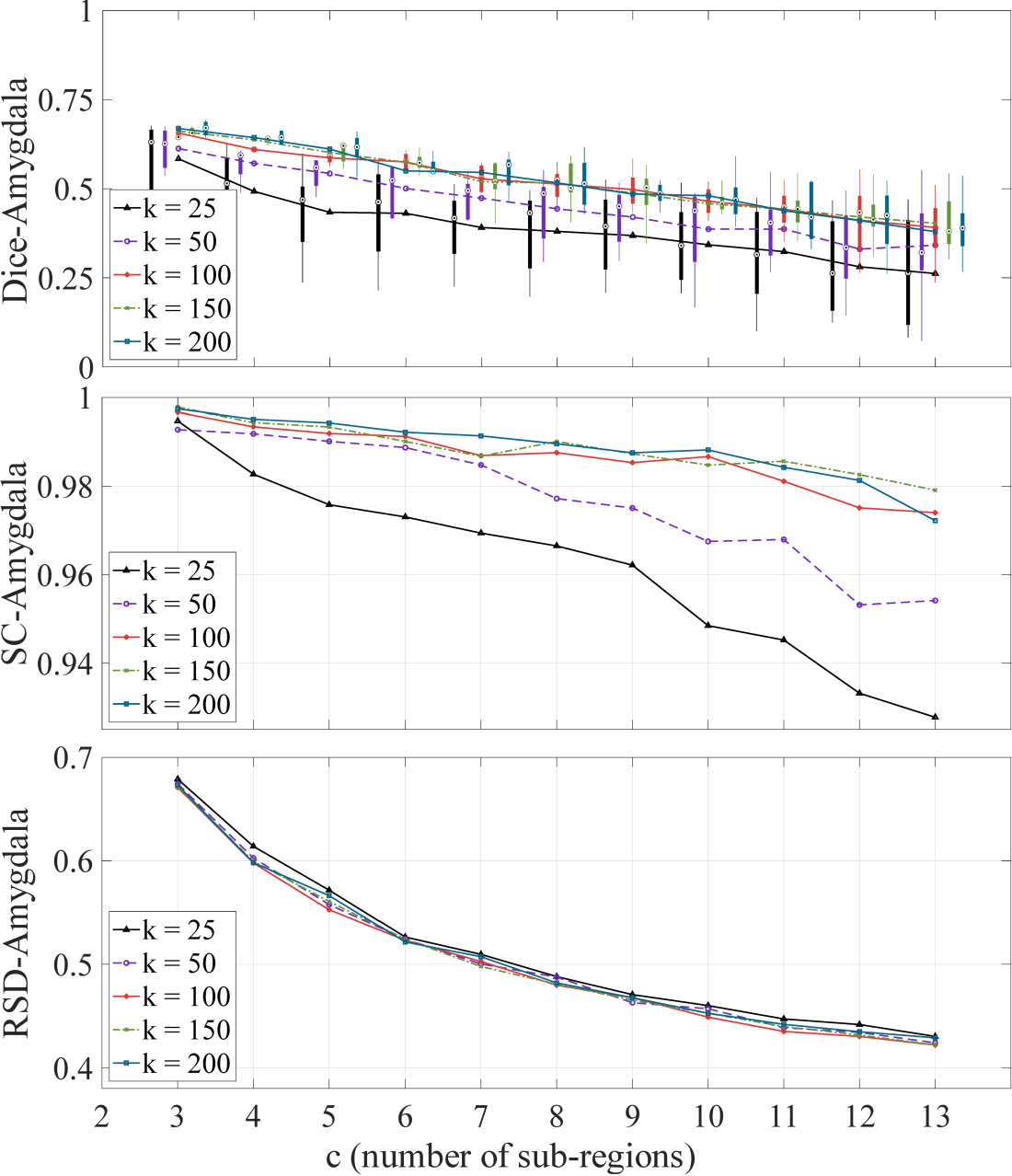}
\caption{Metrics for amygdala parcellation.}\label{Metrics for amygdala parcellation.}
\end{figure}

\begin{figure}[H]
\centering
\includegraphics[width=0.55\linewidth]{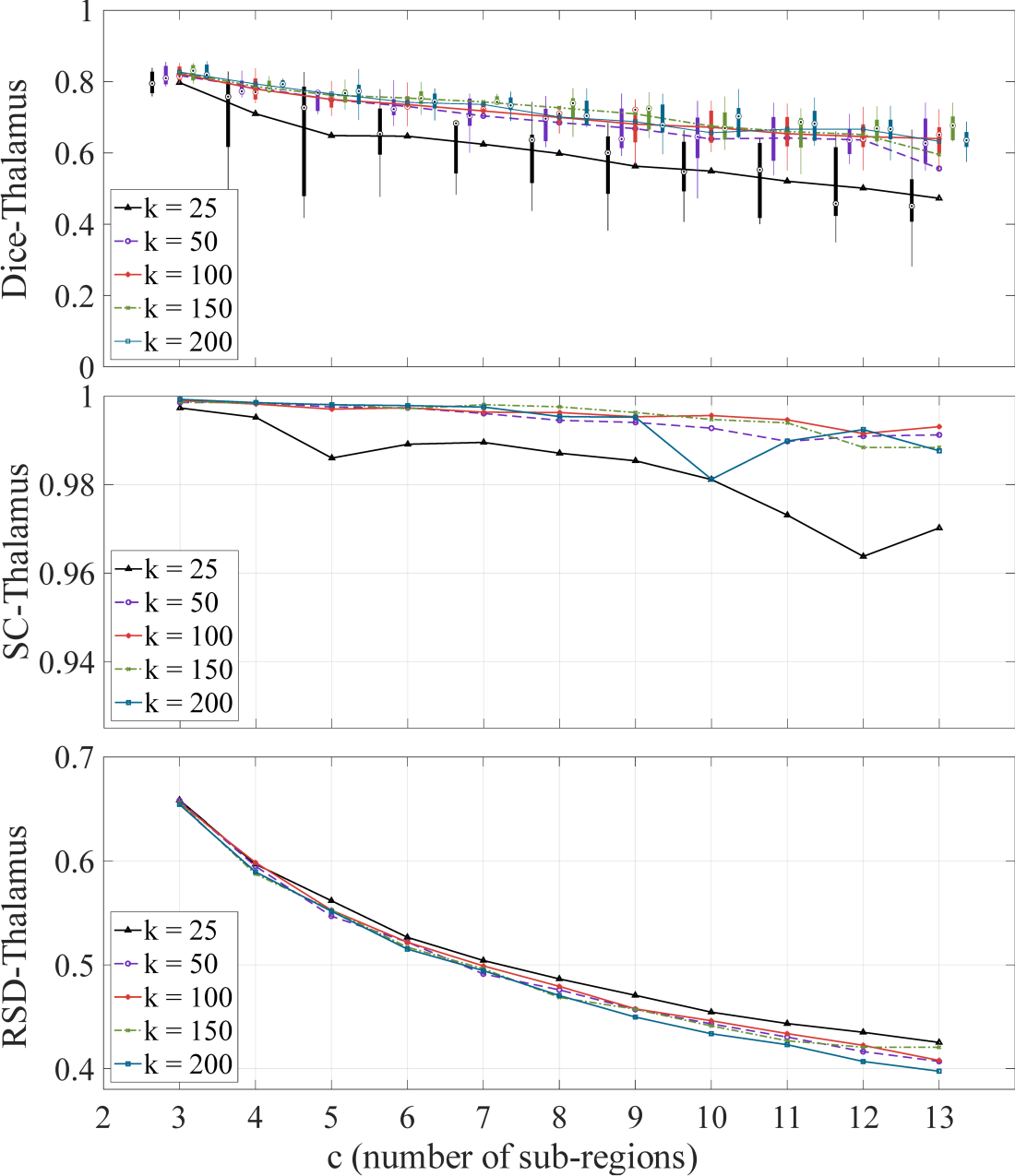}
\caption{Metrics for thalamus parcellation.}\label{Metrics for thalamus parcellation.}
\end{figure}

\begin{figure}[H]
\centering
\includegraphics[width=0.9\linewidth]{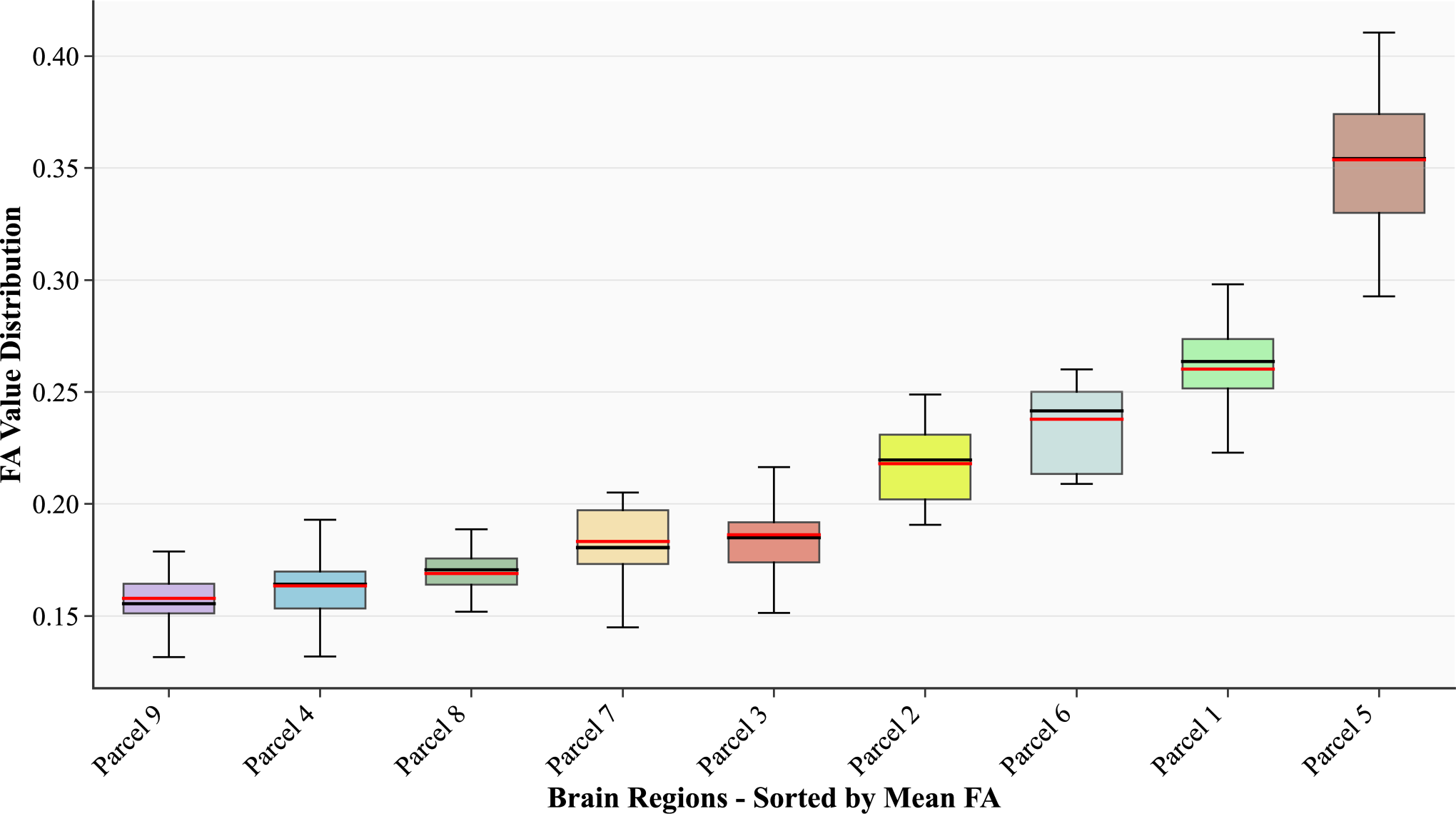}
\caption{FA value distribution across 9 amygdala parcels.}\label{FA value distribution across 9 amygdala parcels.}
\end{figure}

\begin{figure}[H]
\centering
\includegraphics[width=0.9\linewidth]{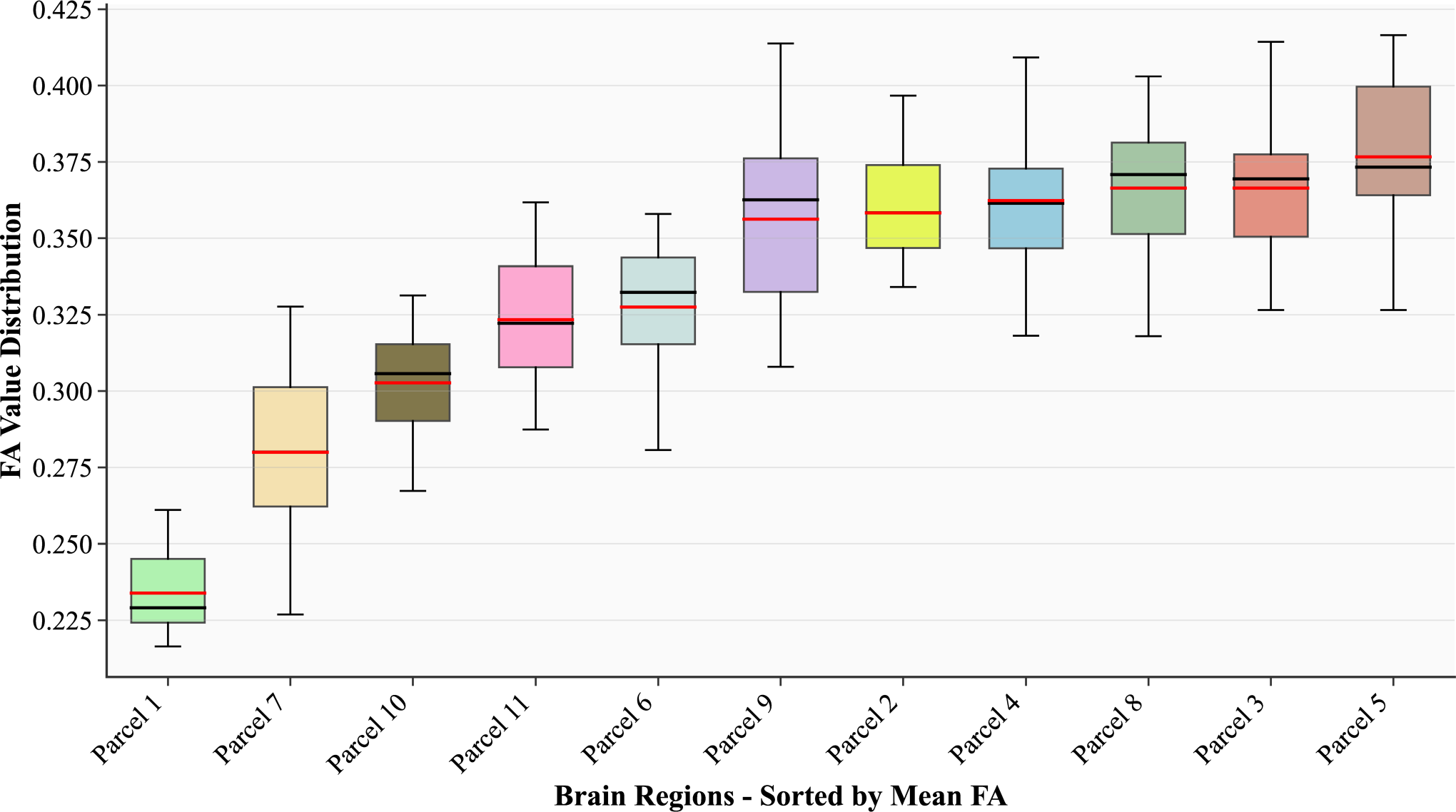}
\caption{FA value distribution across 11 thalamus parcels.}\label{FA value distribution across 11 thalamus parcels.}
\end{figure}

Figures \ref{Metrics for amygdala parcellation.} and \ref{Metrics for thalamus parcellation.} give the metrics obtained for each set of parameter settings for the amygdala and thalamus, respectively.
Regarding the impact of the number of parcels c, we observe that as c increases, the distinctiveness between parcels improves (indicated by lower \textit{RSD} values).
However, this comes at the cost of reduced spatial overlap (lower \textit{Dice} scores) and decreased spatial continuity of the parcels (lower \textit{SC} values).
As for the effect of the number of streamline clusters k, the performance across the three metrics generally improves as k increases gradually, reaching an optimal point under each c setting, after which further increases in k lead to a decline in performance.
We make these choices after comprehensively considering the \textit{Dice}, \textit{SC}, and \textit{RSD} metrics, as well as the additional computational overhead associated with increasing the k value.
We ultimately choose k = 100 and c = 9 for the amygdala, and k = 150 and c = 11 for the thalamus. To validate the biological relevance of our parcellation results, we compute FA values for each parcel within the segmented nucleus structures. For each subject, FA values are calculated for all voxels within each identified parcel, and the distributions of the parcel mean FA are analyzed across subjects. The results are presented as box plots in Figures~\ref{FA value distribution across 9 amygdala parcels.} and~\ref{FA value distribution across 11 thalamus parcels.}. For the amygdala (Figure~\ref{FA value distribution across 9 amygdala parcels.}), FA values range from low anisotropy regions (Parcels 9 and 4, median $\sim$0.16) to high anisotropy regions (Parcel 5, median $\sim$0.35). The thalamus parcellation (Figure~\ref{FA value distribution across 11 thalamus parcels.}) shows similar heterogeneity, with FA values ranging from $\sim$0.24 to $\sim$0.40 across different parcels. These distinct FA distributions indicate that our parcellation captures meaningful microstructural differences between subdivisions.

\subsection{Comparison across Different Methods}
\label{Comparison across Different Methods}

We compare the following methods:
1) the baseline k-means method that applies the traditional k-means clustering method directly on the input connectivity feature representation,
2) the feat-orig method that applies the proposed network on the original feature vector \cite{Xu2023} (as introduced in Section \ref{Feature Extraction via Streamline Cluster Dilation and Smoothing}) without the streamline cluster dilation and smoothing process,
3) the net-orig method that applies the original deep clustering network \cite{Yang2017} (as introduced in Section \ref{Fine-Scale Nucleus Parcellation via Deep Clustering}) on our proposed connectivity feature representation, and
4) our method that applies the proposed autoencoder-based k-means-friendly network (Section \ref{Fine-Scale Nucleus Parcellation via Deep Clustering}) to the proposed connectivity feature representation (Section \ref{Voxel-wise Connectivity Feature Representation}).
Table \ref{Quantitative comparison across different methods for the amygdala parcellation.} and Table \ref{Quantitative comparison across different methods for the thalamus parcellation.} give the comparison results, where our method in general obtains the best performance across all compared methods.


\begin{table}[H]
    \centering
    \caption{Quantitative comparison across different methods for the amygdala parcellation. For the Dice overlap, the averaged metric for each parcel $p$ is also provided.}
    \renewcommand{\arraystretch}{1.0}
    \setlength{\tabcolsep}{2pt}
    \begin{tabular}{lcccccccccccc}
        \toprule
        \multirow{2}{*}{\textbf{Amygdala}} & \multirow{2}{*}{\textit{SC}} & \multirow{2}{*}{\textit{RSD}} & \multicolumn{10}{c}{\textit{Dice}} \\
        \cmidrule(lr){4-13}
        & & & Avg & $p_1$ & $p_2$ & $p_3$ & $p_4$ & $p_5$ & $p_6$ & $p_7$ & $p_8$ & $p_9$ \\
        \midrule
        \textit{k-means}      & 0.57 & 0.48 & 0.19 & 0.22 & 0.37 & 0.16 & 0.23 & 0.15 & 0.10 & 0.25 & 0.10 & 0.17 \\
        \textit{feat-orig}    & 0.75 & 0.48 & 0.20 & 0.17 & 0.60 & 0.12 & 0.28 & 0.10 & 0.09 & 0.16 & 0.12 & 0.14 \\
        \textit{net-orig}     & 0.98 & \textbf{0.46} & 0.42 & 0.33 & 0.50 & 0.44 & 0.34 & 0.48 & 0.54 & 0.44 & 0.43 & 0.30 \\
        \textbf{ours}         & \textbf{0.99} & 0.47 & \textbf{0.50} & 0.58 & 0.50 & 0.43 & 0.58 & 0.46 & 0.50 & 0.51 & 0.46 & 0.45 \\
        \bottomrule
    \end{tabular}
    \label{Quantitative comparison across different methods for the amygdala parcellation.}
\end{table}


\begin{table}[H]
    \centering
    \caption{Quantitative comparison across different methods for the thalamus parcellation. For the Dice overlap, the averaged metric for each parcel $p$ is also provided.}
    \renewcommand{\arraystretch}{1.0}
    \setlength{\tabcolsep}{1.8pt}
    \begin{tabular}{lcccccccccccccc}
        \toprule
        \multirow{2}{*}{\textbf{Thalamus}} & \multirow{2}{*}{\textit{SC}} & \multirow{2}{*}{\textit{RSD}} & \multicolumn{12}{c}{\textit{Dice}} \\
        \cmidrule(lr){4-15}
        & & & Avg & $p_1$ & $p_2$ & $p_3$ & $p_4$ & $p_5$ & $p_6$ & $p_7$ & $p_8$ & $p_9$ & $p_{10}$ & $p_{11}$ \\
        \midrule
        \textit{k-means}      & 0.35 & 0.44 & 0.20 & 0.29 & 0.19 & 0.16 & 0.12 & 0.33 & 0.14 & 0.36 & 0.27 & 0.13 & 0.12 & 0.14 \\
        \textit{feat-orig}    & 0.55 & 0.43 & 0.21 & 0.53 & 0.18 & 0.28 & 0.22 & 0.10 & 0.10 & 0.11 & 0.41 & 0.04 & 0.24 & 0.10 \\
        \textit{net-orig}     & 0.99 & 0.42 & 0.65 & 0.66 & 0.68 & 0.68 & 0.56 & 0.69 & 0.67 & 0.61 & 0.65 & 0.59 & 0.65 & 0.68 \\
        \textbf{ours}         & \textbf{0.99} & \textbf{0.42} & \textbf{0.66} & 0.64 & 0.72 & 0.73 & 0.58 & 0.67 & 0.67 & 0.58 & 0.68 & 0.60 & 0.72 & 0.67 \\
        \bottomrule
    \end{tabular}
    \label{Quantitative comparison across different methods for the thalamus parcellation.}
\end{table}

\subsection{Visualization of Parcellation Results and Comparison with Existing \mbox{Atlas}}
\label{Visualization of Parcellation Results and Comparison with Existing Atlas}

\begin{figure}[H]
\centering
\includegraphics[width=1\linewidth]{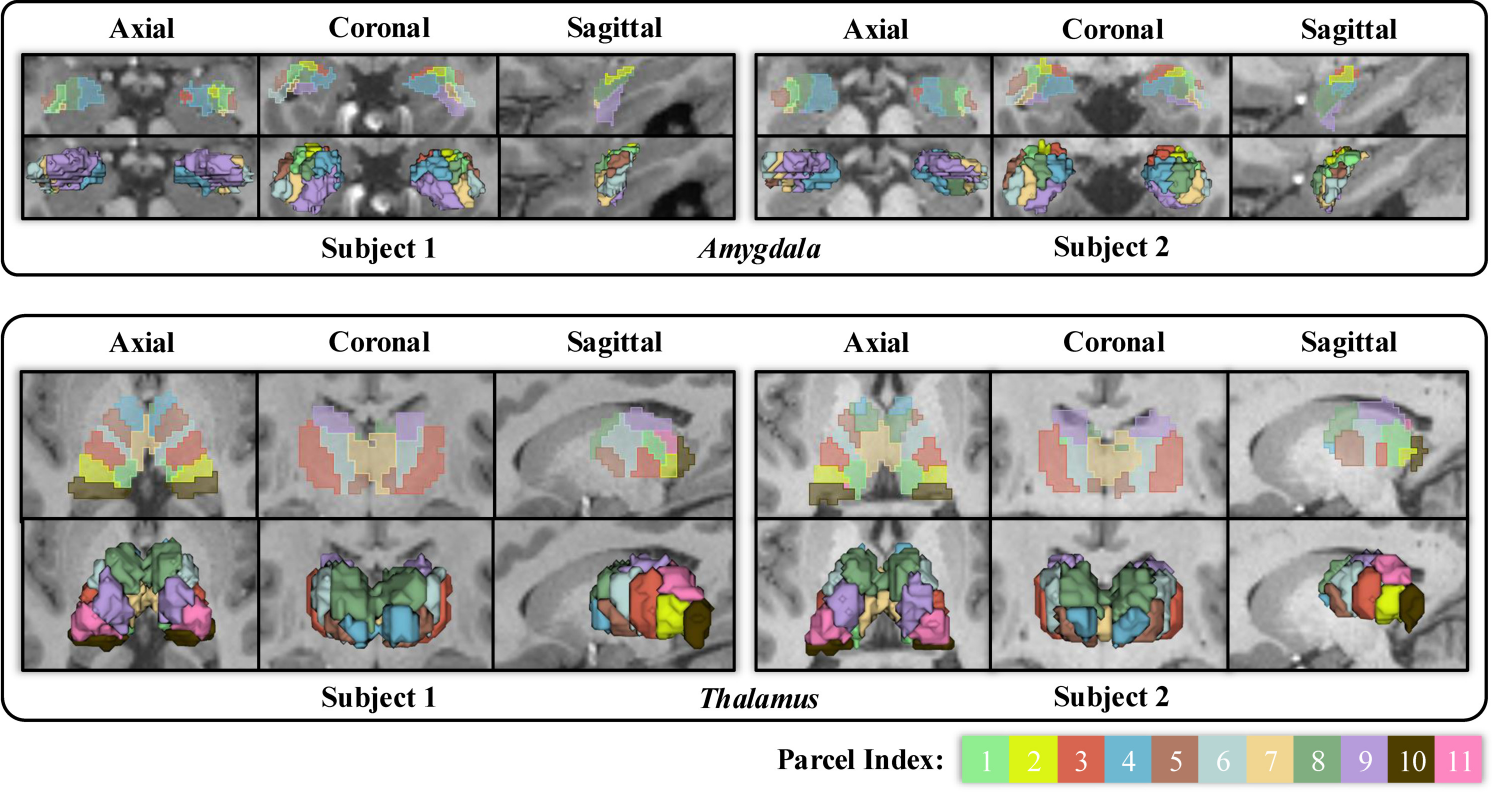}
\caption{Visualization of the parcellation results obtained using the proposed method of two randomly selected subjects.}\label{Visualization of the parcellation results of two randomly selected subjects.}
\end{figure}

To assess our parcellation methodology, we conduct a comparative analysis with established atlases.
Specifically, for the amygdala parcellation, we leverage the SPM Anatomy Toolbox \cite{Eickhoff2005}, which categorizes the amygdala into three distinct subdivisions: LB, CM, and SF segments.
For the thalamus parcellation, we utilize the Melbourne Subcortex Atlas s3 \cite{Tian2020}, which classifies the thalamus into six delineated regions: medial ventroposterior (VPm), lateral ventroposterior (VPl), inferior ventroanterior (VAi), superior ventroanterior (VAs), medial dorsoanterior (DAm), and lateral dorsoanterior (DAl). 

\begin{figure}[H]
\centering
\includegraphics[width=0.8\linewidth]{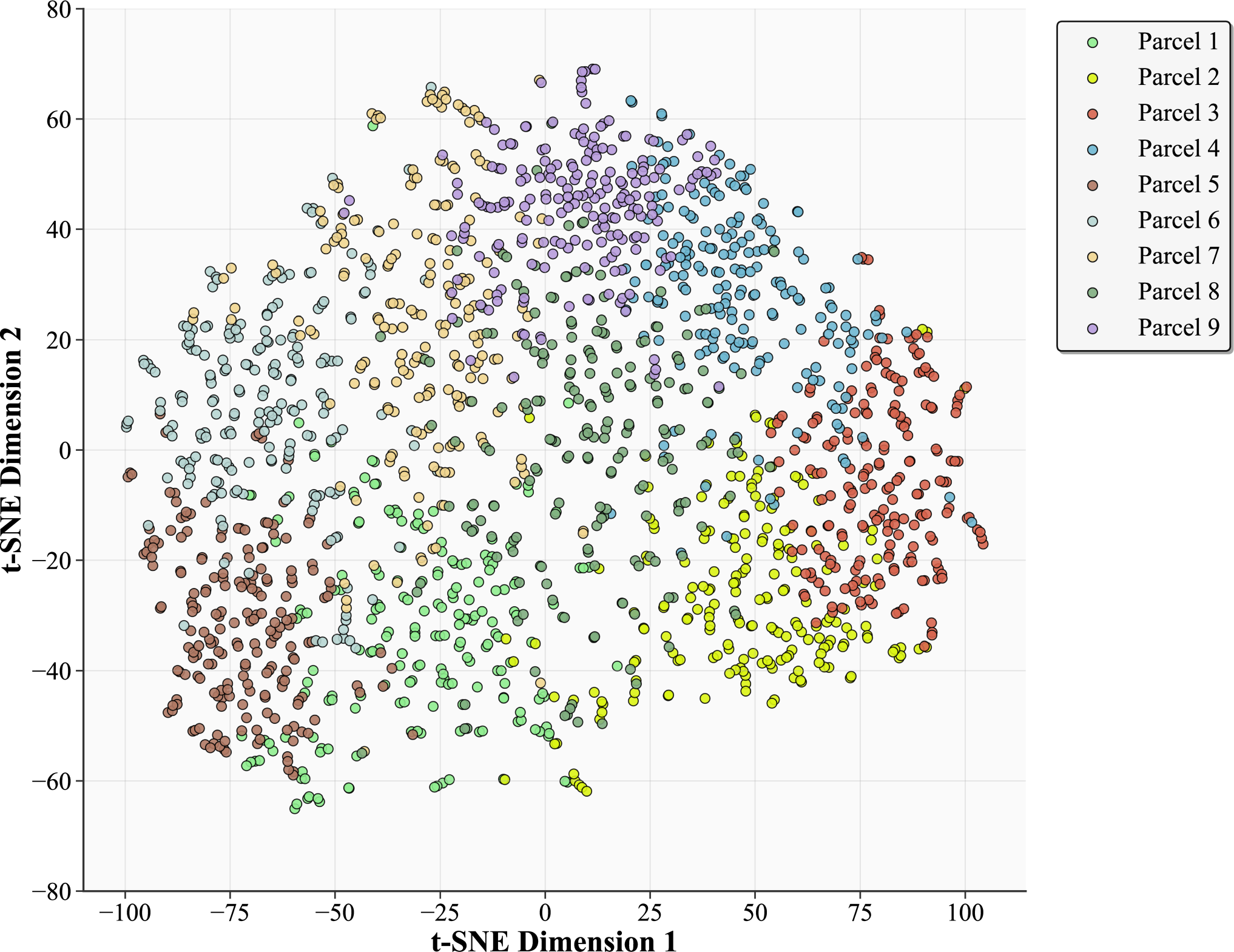}
\caption{t-SNE visualization of learned voxel embeddings for amygdala parcellation.}\label{t-SNE visualization of learned voxel embeddings for amygdala parcellation.}
\end{figure}

\begin{figure}[H]
\centering
\includegraphics[width=0.8\linewidth]{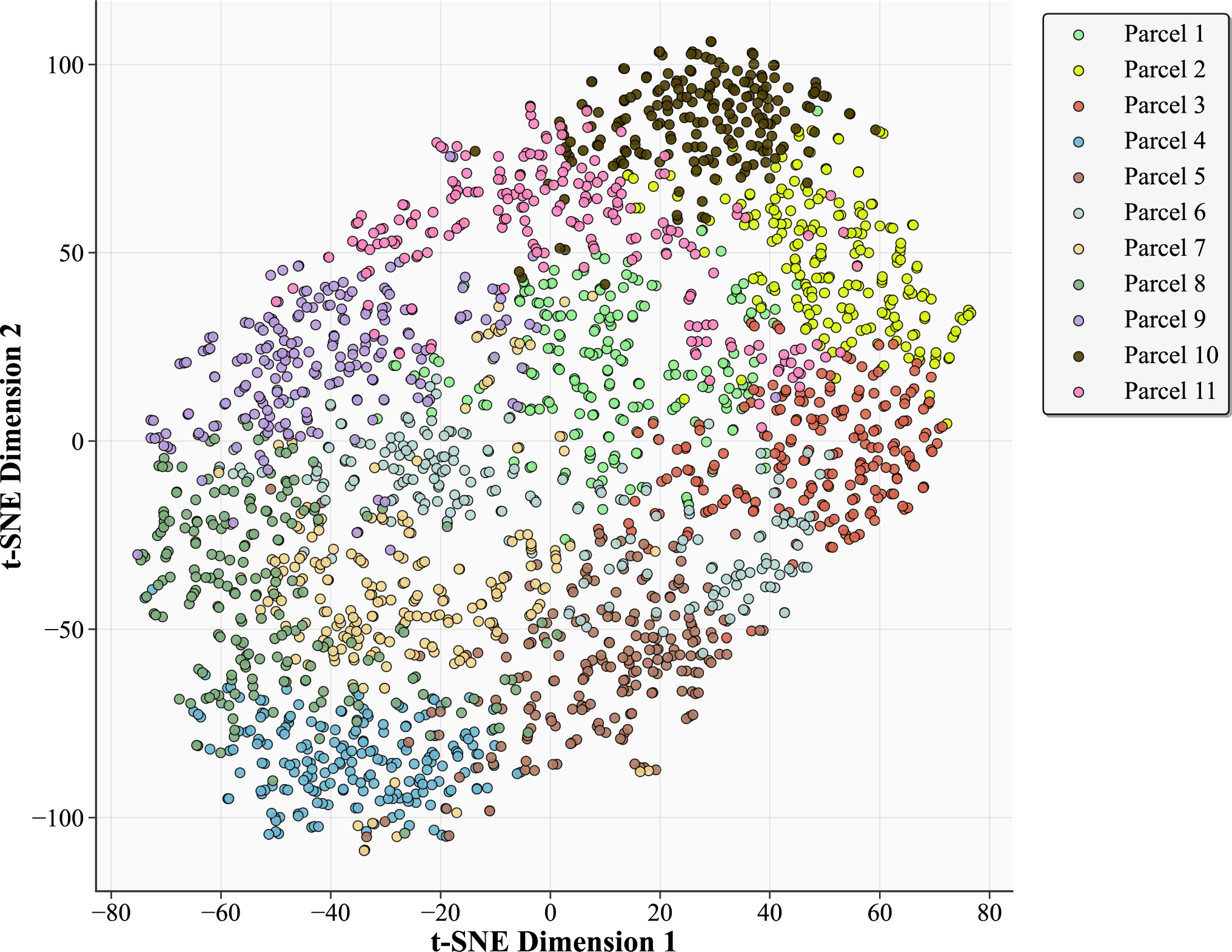}
\caption{t-SNE visualization of learned voxel embeddings for thalamus parcellation.}\label{t-SNE visualization of learned voxel embeddings for thalamus parcellation.}
\end{figure}

To visualize the quality of the learned latent representations, we perform t-SNE \cite{Maaten2008} analysis on the autoencoder embeddings. For each parcel, we randomly sample 10 voxels per subject and visualize their embeddings in the learned feature space (Figures \ref{t-SNE visualization of learned voxel embeddings for amygdala parcellation.} and \ref{t-SNE visualization of learned voxel embeddings for thalamus parcellation.}). The t-SNE plots demonstrate clear clustering of voxels according to their parcel assignments, confirming that our autoencoder successfully learns discriminative features that align well with the final parcellation results.

\begin{figure}[H]
\centering
\includegraphics[width=1\linewidth]{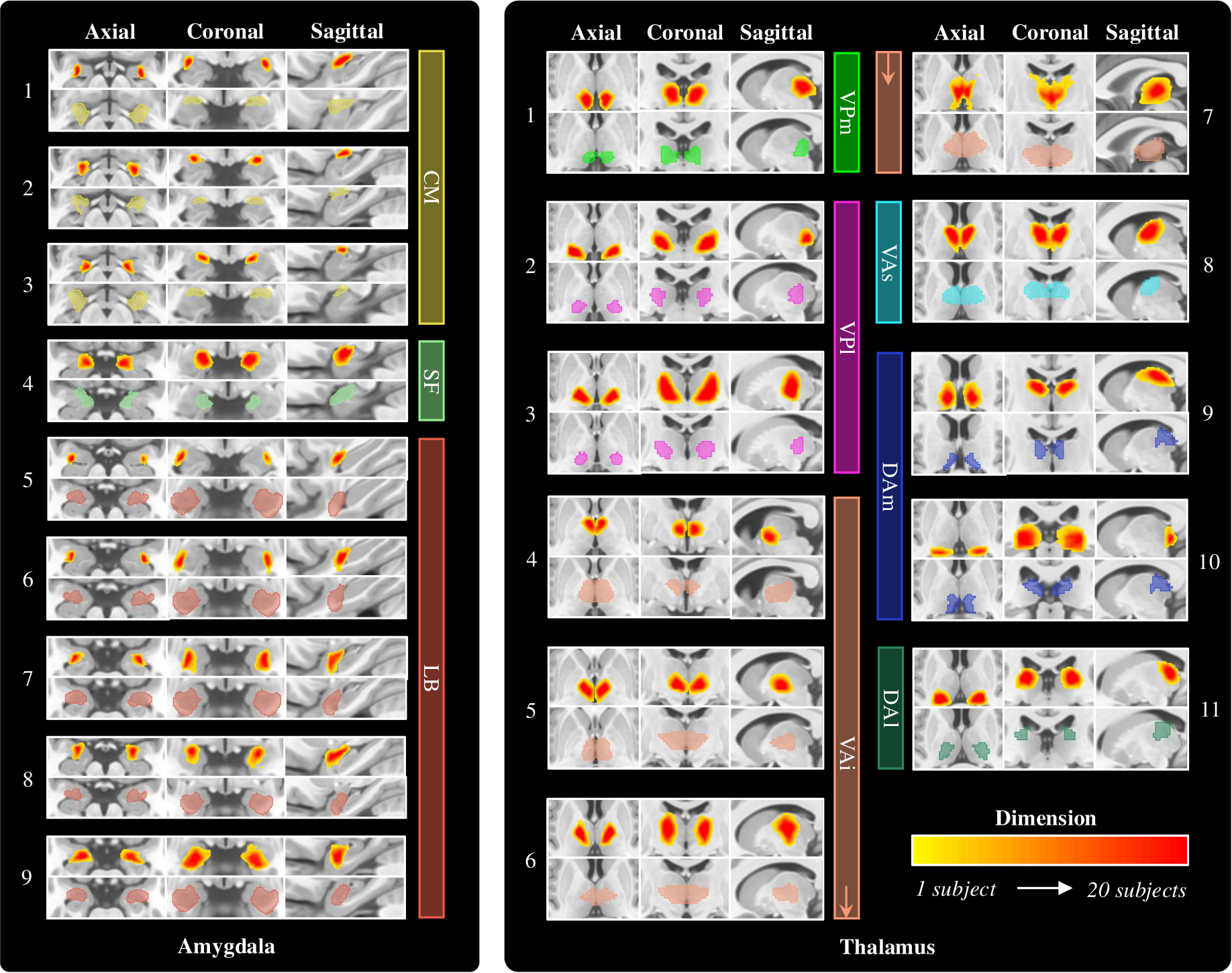}
\caption{Visualization of the group-wise parcellation result (the heatmap presented in the top row for each parcel), with comparison to the existing parcellation atlases (the color label presented in the top row for each parcel).}\label{Visualization of the group-wise parcellation result, with comparison to the existing parcellation atlases.}
\end{figure}

Figure \ref{Visualization of the parcellation results of two randomly selected subjects.} and Figure \ref{Visualization of the group-wise parcellation result, with comparison to the existing parcellation atlases.} give a visualization of our subject-specific and group-wise parcellations of the amygdala and thalamus, respectively, as well as the SPM atlas for the amygdala and the Melbourne Subcortex Atlas s3 for the thalamus.
For the amygdala, we have three parcels for CM, two for SF, and four for LB, each highly visually overlapping with the corresponding atlas parcel.
For the thalamus, we have one parcel for VPm, two for VPl, four for VAi, one for VAs, two for DAm, and one for DAl.
Quantitatively, the average Dice coefficient between our parcels and the corresponding atlas parcels is 0.72 for the amygdala and 0.62 for the thalamus.

\section{Discussion and Conclusion}
\label{Discussion and Conclusion}

In this work, we propose \textit{DeepNuParc}, a novel deep clustering pipeline to perform parcellation of the brain nuclei with dMRI tractography.
In our proposed framework, we first compute a novel voxel-wise connectivity feature representation, with a feature refinement process using the newly proposed streamline cluster dilation and smoothing.
Next, we design an adaptive k-means-friendly autoencoder framework that can compress the feature representation and jointly train with the downstream clustering algorithm.
Finally, we achieve fine-scale parcellation of the brain nucleus structure by clustering voxels into different groups.

A crucial step in our \textit{DeepNuParc} method involves the explicit reconstruction of dMRI tractography streamlines to create the feature representation, as opposed to relying on connectivity probabilities derived from probabilistic tractography \cite{Avecillas2023, Saygin2011}.
The explicit reconstruction of tractography streamlines allows us to form streamlines that pass through a nucleus into streamline clusters, thereby enriching the data representation.
Furthermore, by constructing each voxel's feature vector based on its traversal by streamline clusters, we transform a complex high-dimensional brain parcellation problem into a simpler one-dimensional vector clustering problem.
This approach enhances both the simplicity and robustness of the algorithm.

Our algorithm has good flexibility and compatibility to be modified and customized.
The entire processing framework is composed of multiple steps connected in sequence, with each step being a module that performs a specific function.
This means we can adjust the hyperparameters of specific modules (e.g., the extent of cluster dilation or the number of streamline clusters) or optimize the algorithms of certain modules (e.g., direct extraction of the nucleus structure) to achieve adjustments or improvements in the results.
Secondly, our algorithm demonstrates a high robustness to parameter choices.
Regardless of how we adjust the parameters k or c, we consistently obtain good metrics.
Specifically, for spatial continuity, we do not explicitly enforce that all voxels in a parcel must be connected, yet over 99\% of the voxels in the same parcel ended up being connected through the neural network's own representation learning.
This indicates that our algorithm effectively utilizes the information present in the data.

We show that our parcellation results are anatomically reasonable when compared to the existing atlases.
For the amygdala, we identify nine different parcels, which show a strong match with the previous atlas of three parcels (CM, SF, LB).
There are three parcels for CM, two for SF, and four for LB.
For the thalamus, we identify eleven parcels, which also match well with the previous atlas of six parcels (VPm, VPl, VAi, VAs, DAm, DAl).
There is one parcel for VPm, two for VPl, four for VAi, one for VAs, two for DAm, and one for DAl.
Each of these parcellated parcels shows a highly visually overlapping with the corresponding atlas parcel.
Overall, our parcellation results, which provide finer parcellation granularity, show good alignment with prior segmentation studies when compared to existing atlases.
This suggests that our algorithm holds promise for exploring subregions of brain nucleus structures.

Potential limitations of the present study, including suggested future work to address limitations, are as follows.
First, in our current work, we apply \textit{DeepNuParc} to study the amygdala and thalamus.
However, our method can be generally applied to any other brain nucleus structures or gray matter structures that are connected with white matter tracts.
Therefore, it would be of interest in future work to test how our method works on other anatomical structures.
Second, our current experiments are performed using the HCP dataset acquired from young healthy adults.
Future work should involve training and testing the model on datasets from diverse sources and populations, though this presents a significant challenge.  For example, clinical data with lower resolution (typically 2-3mm), fewer diffusion directions, and lower b-values may present challenges for tractography quality and parcellation performance. Individuals of different ages and those with neurological conditions may exhibit anatomical variability and altered white matter integrity that could affect connectivity patterns.
Third, our current method focuses on using structural connectivity to perform the parcellation, while it is an interesting direction to integrate multimodal data such as fMRI to improve the functional coherence of the results. Fourth, in our current study, we validate the biological relevance of the obtained parcels by comparing them with existing atlases and conducting FA distribution and embedding analyses. However, the underlying biological significance of each parcel remains unclear and warrants further investigation using anatomical techniques such as tracer studies. Fifth, to balance computational efficiency, we used data from 100 HCP subjects for model training and testing. While this sample size has proven sufficient for dMRI-based machine/deep learning tasks in our prior work \cite{ZhangDDMReg2021, ZhangAtlas2018, ZhangSegmentation2021, ZhangDeepWMA2020}, we expect that increasing it would further enhance our method's performance.

Overall, we show the ability to use dMRI tractography and deep learning to perform fine-scale parcellation of brain nuclei.
Our method may provide a useful tool to explore insights into the detailed structure and function of brain nuclei.

\section{Ethics Statement}
\label{Ethics Statement}

This study was conducted retrospectively using public HCP imaging data. No ethical approval was required.

\section{Acknowledgements}

This work is in part supported by the National Key R\&D Program of China (No. 2023YFE0118600), and the National Natural Science Foundation of China (No. 62371107).

\bibliographystyle{unsrt}
\bibliography{references_clean}

\end{document}